\documentclass[11pt]{article}

\usepackage{graphicx}
\usepackage{a4wide}
\usepackage{subfig}
\usepackage{amsfonts}

\newcommand{\nit}{\noindent}

\newcommand{\np}{\newpage}
\newcommand{\dsp}{\displaystyle}
\newcommand{\vs}[1]{\vspace{#1 ex}}
\newcommand{\hs}[1]{\hspace{#1 em}}
\newcommand{\bfr}{\begin{flushright}}
\newcommand{\efr}{\end{flushright}}
\newcommand{\bc}{\begin{center}}
\newcommand{\ec}{\end{center}}
\newcommand{\ben}{\begin{enumerate}}
\newcommand{\een}{\end{enumerate}}

\newcommand{\be}{\begin{equation}}
\newcommand{\ee}{\end{equation}}
\newcommand{\ba}{\begin{array}}
\newcommand{\ea}{\end{array}}
\newcommand{\ct}{\cite}
\newcommand{\bit}{\bibitem}

\newcommand{\del}{\delta}

\newcommand{\ve}{\varepsilon}

\newcommand{\thg}{\theta}
\newcommand{\kg}{\kappa}
\newcommand{\lb}{\lambda}
\newcommand{\sg}{\sigma}

\newcommand{\vf}{\varphi}
\newcommand{\og}{\omega}

\newcommand{\Del}{\Delta}

\newcommand{\lh}{\left(}
\newcommand{\rh}{\right)}
\newcommand{\ld}{\left.}

\newcommand{\der}{\partial}

\begin{document}

\pagestyle{empty} 

\hfill NIKHEF/2011-007
\vs{7}

\bc
{\Large{\bf Geodesic deviations: }} \\
\vs{2}

{\Large{\bf modeling extreme mass-ratio systems}}\\
\vs{2}

{\Large{\bf  and their gravitational waves}}
\vs{7}

{\large{G.\ Koekoek$^1$ and J.W.\ van Holten$^2$}}
\vs{2}

{\large{Nikhef, Amsterdam}}
\vs{3}

\today
\vs{10}

{\small{\bf Abstract}}
\ec

\nit
{\small
The method of geodesic deviations has been applied to derive accurate analytic approximations to geodesics in 
Schwarzschild space-time. The results are used to construct analytic expressions for the source terms in the 
Regge-Wheeler and Zerilli-Moncrief equations, which describe the propagation of gravitational waves emitted by 
a compact massive object moving in the Schwarzschild background space-time. The wave equations are  
solved numerically to provide the asymptotic form of the wave at large distances for a series of non-circular 
bound orbits with periastron distances up to the ISCO radius, and the power emitted in gravitational waves by 
the extreme-mass ratio binary system is computed. The results compare well with those of purely numerical 
approaches.}

\vfill
\footnoterule 
\nit
{\footnotesize{$^1$ e-mail: gkoekoek@nikhef.nl } \\ $^2$ email: v.holten@nikhef.nl}

\np
\pagestyle{plain}
\pagenumbering{arabic}

\section{Introduction \label{intro}}

In a recent paper \ct{GKJWvH} we have implemented a new kind of perturbation theory for geodesics in curved space-time,
an improved version of the method of geodesic deviations \ct{BalHolKer,OriginalEpicycle,jwvh,ker-col}. In contrast to the 
well-established post-newtonian scheme \ct{Maggiore,PNoverview}, this perturbation theory is fully relativistic by construction, 
and is expected to become especially relevant for motion in strongly curved space-time regions. This holds in particular for 
matter moving in the vicinity of black-hole horizons. Indeed, our method seems particularly well-suited to describe extreme 
mass-ratio binary systems, formed by a compact object --e.g., a neutron star or stellar-mass black hole-- orbiting a giant 
black hole such as found in the center of many galaxies, with masses exceeding a million solar masses and sometimes 
much more. 

Like the well-known compact binary systems of neutron stars \ct{ht,reviewPSR1913}, these extreme mass-ratio binaries 
are expected to lose energy by the emission of gravitational radiation. Discovery and analysis of this radiation would be
a direct way to probe the strong-curvature region around the giant black hole. In this paper we summarize our covariant 
perturbation scheme and apply it to the calculation of gravitational waves emitted by extreme-mass ratio binaries. 
We will do this for the rotationally invariant case of Schwarzschild geometry, describing a non-rotating black hole.
Gravitational waves are represented by the linear metric perturbations propagating on this background geometry, in 
particular the perturbations created by a test mass orbiting the black hole. Such perturbations are described by the 
Regge-Wheeler and Zerilli-Moncrief equations \ct{RW,zerilli, moncrief}, the solutions of which represent the two physical 
modes of gravitational waves in a Schwarzschild background. 

Our perturbative solution for the geodesics allows us to write the source terms of the Regge-Wheeler-Zerilli-Moncrief
equations in analytical form. The equations can then be solved numerically using the algorithm of Lousto and Price 
\ct{lousto-price, Martel2, Martel}. Good agreement with the results existing in the literature is obtained. By just varying 
the initial data many solutions can be found in a very efficient way. 

\section{Parametrization of orbits \label{GeodesicDeviationMethod}}

Complete exact solutions for the geodesic equations in Schwarzschild space-time are known for special cases, 
including circular orbits, the straight plunge and special inspiraling motions \ct{chandra,MTW}. More general orbits can 
be constructed in a covariant scheme as perturbations of these special analytic solutions; in this paper we consider 
especially bound orbits as covariant deformations of circular ones, using the results obtained in \ct{GKJWvH}. 

The orbits of test masses are described by geodesics $x^{\mu}(\tau)$ parametrized by the proper time $\tau$. We 
choose standard Schwarzschild-Droste co-ordinates to parametrize the line element in the form
\be
d\tau^2 = \lh 1 - \frac{GM}{r} \rh dt^2 - \frac{dr^2}{1 - \frac{2M}{r}} - r^2 d\thg^2 - r^2 \sin^2 \thg\, d\vf^2.
\label{2.0}
\ee
Moreover, as the conservation of angular momentum guarantees all orbits to be planar, we can
fix the plane of the orbit to be the equatorial plane $\thg = \pi/2$. With initial conditions $t(0) = \vf(0) = 0$, circular orbits 
are then given by 
\be
t(\tau) = \frac{\tau}{\sqrt{1 - \frac{3M}{R}}}, \hs{2} r = R, \hs{2} 
\vf(\tau) = \sqrt{\frac{M}{R^3}}\, \frac{\tau}{\sqrt{1 - \frac{3M}{R}}},
\label{2.1}
\ee
where $R$ is the constant radial co-ordinate of the motion. Circular geodesics are characterized by special values of the
constants of motion $(\ve, \ell)$ corresponding to energy and angular momentum per unit of test mass:
\be
\ve \equiv \lh 1 - \frac{2M}{r} \rh \frac{dt}{d\tau} = \frac{1 - \frac{2M}{R}}{\sqrt{1 - \frac{3M}{R}}}, \hs{2} 
\ell \equiv r^2\, \frac{d\vf}{d\tau} = \sqrt{\frac{MR}{1 - \frac{3M}{R}}}.
\label{2.2}
\ee
For more general orbits $\ve$ and $\ell$ are unrelated, and $r$ is not constant. Using the same initial conditions
bound orbits can be parametrized as deformations of  circular orbits of the form
\be 
\ba{lll}
t(\tau) & = & \dsp{ a^t_0 \tau + \sum_{n = 1}^{\infty} a_n^t \sin n \og \tau, }\\
 & & \\
r(\tau) & = & \dsp{ a^r_0 + \sum_{n = 1}^{\infty} a^r_n \cos n \og \tau, }\\
 & & \\ 
\vf(\tau) & = & \dsp{ a^{\vf}_0 \tau + \sum_{n=1}^{\infty} a^{\vf}_n \sin n \og \tau, }
\ea
\label{2.3}
\ee
where both the amplitudes $a^{\mu}_n$ and the angular frequency $\og$ are to be expanded as power series 
in terms of a deformation parameter $\sg$, the series for the co-efficients $a_n^{\mu}$ starting at order $\sg^n$. 
The parameter $\sg$ is defined in terms of the distance between the perturbed and the circular orbit at $\tau = 0$, 
\be 
d\sg^2 = g_{\mu\nu} \ld dx^{\mu} dx^{\nu} \right|_{\tau = 0}. 
\label{2.4}
\ee
Analytic approximations to the orbits are obtained by cutting the expansions at some finite order in $\sg$: 
\be
\ba{lll}
\dsp{ a_0^t = \frac{1}{\sqrt{1 - \frac{3M}{R}}} + \sg v_1^t + \frac{1}{2} \sg^2 v_2^t + ..., }& \dsp{
a_1^t = \sg n_1^t + \frac{1}{2} \sg^2 n_2^t + ..., }& \dsp{ a_2^t = \frac{1}{2} \sg^2 m_2^t + ..., }\\
\\
\dsp{ a_0^r = R + \sg \Del_1 + \frac{1}{2}\, \sg^2 \Del_2 + ..., }& \dsp{
a_1^r = \sg n_1^r + \frac{1}{2}\, \sg^2 n_2^r + ..., }& \dsp{ a_2^r = \frac{1}{2}\, \sg^2 m_2^r + ..., }\\
\\
\dsp{ a_0^{\vf} = \sqrt{\frac{M}{R^3}} \frac{1}{\sqrt{1 - \frac{3M}{R}}} + \sg v_1^{\vf} + \frac{1}{2} \sg^2 v_2^{\vf} + ..., }& 
\dsp{ a_1^{\vf} = \sg n_1^{\vf} + \frac{1}{2} \sg^2 n_2^{\vf} + ..., }& \dsp{ a_2^{\vf} = \frac{1}{2} \sg^2 m_2^{\vf} + ..., }
\ea
\label{2.5.1}
\ee
with coefficients $a_n^{\mu}$, $n\geq3$, and the dots representing terms of order $\sg^3$ and higher.  
Also the angular frequency is to be computed order by order in the expansion parameter:
\be
\og = \og_0 + \sg \og_1 + ...
\label{2.5.2}
\ee
The explicit expressions for the various terms above are given by \\
a.\ for the time co-ordinate:
\be
\ba{l}
\dsp{ v_1^t = -  \Del_1\, \frac{3 M}{2R^2}\, \frac{1}{\lh 1 - \frac{3M}{R} \rh^{3/2}}
 \equiv - \nu \Del_1, }\\
 \\
\dsp{ v_2^t = - \nu \lh \Del_2 + \frac{\Del_1^2}{2R}\, \frac{2 - \frac{36M}{R} + \frac{153M^2}{R^2} - \frac{162M^3}{R^3}}{
 \lh 1 - \frac{3M}{R} \rh \lh 1 - \frac{6M}{R} \rh} - \frac{n_1^{r\, 2}}{R}\, \frac{1 + \frac{M}{R}}{1 - \frac{2M}{R}} \rh,    }
\ea
\label{2.5.3}
\ee
and
\be
\ba{l}
\dsp{ n_1^t = - 2n_1^r\, \sqrt{\frac{M}{R}}\, \frac{1}{\lh 1 - \frac{2M}{R} \rh \sqrt{1 - \frac{6M}{R}} } \equiv - \lb n_1^r , }\\
 \\
\dsp{ n_2^t = - \lb\, \frac{n_1^r \Del_1}{R}\, \frac{3 - \frac{34M}{R} + \frac{72M^2}{R^2}}{\lh 5 - \frac{18M}{R} \rh
 \lh 1 - \frac{2M}{R} \rh}, }\\
 \\
\dsp{ m_2^t =  \lb\, \frac{n_1^{r\,2}}{2 R}\, \frac{2 - \frac{15M}{R} + \frac{14M^2}{R^2}}{\lh 1 - \frac{2M}{R} \rh
 \lh 1 - \frac{6M}{R} \rh}; }
\ea
\label{2.5.4}
\ee
b.\ for the radial co-ordinate:
\be 
\ba{l}
\dsp{ \Del_1 + n_1^r = \sqrt{1 - \frac{2M}{R}}, }\\ 
 \\
\dsp{ n_2^r = \frac{4n_1^r \Del_1}{R}\, \frac{2 - \frac{15M}{R} + \frac{30 M^2}{R^2}}{\lh 5 - \frac{18M}{R} \rh 
 \lh 1 - \frac{2M}{R} \rh \lh 1 - \frac{6M}{R} \rh}, }\\
 \\
\dsp{ m_2^r =  - \frac{n_1^{r\,2}}{R}\, \frac{1 - \frac{7M}{R}}{1 - \frac{6M}{R}}, }
\ea
\label{2.5.6}
\ee
with $\Del_{n}$, $n = 1,2,...$, free parameters; \\
c.\ for the angular co-ordinate:
\be
\ba{l}
\dsp{ v_1^{\vf} = - \frac{3\Del_1}{2R^2}\, \sqrt{\frac{M}{R}}\, \frac{1 - \frac{2M}{R}}{\lh 1 - \frac{3M}{R} \rh^{3/2}} 
 \equiv - \kg \Del_1, }\\
 \\
\dsp{ v_2^{\vf} = - \kg \lh \Del_2 + \frac{\Del_1^2}{2R} \frac{1 - \frac{25M}{R} + \frac{153M^2}{R^2} - \frac{360M^3}{R^3}
 + \frac{324M^4}{R^4}}{\lh 1 - \frac{2M}{R} \rh \lh 1 - \frac{3M}{R} \rh \lh 1 - \frac{6M}{R} \rh} 
 - \frac{n_1^{r\,2}}{R}\, \frac{1 + \frac{M}{R}}{1 - \frac{2M}{R}} \rh,    }
\ea
\label{2.5.7}
\ee
and
\be
\ba{l}
\dsp{ n_1^{\vf} =  - \frac{2n_1^r}{R}\, \frac{1}{\sqrt{ 1 - \frac{6M}{R} }} \equiv - \mu n_1^r, }\\
 \\
\dsp{ n_2^{\vf} =   \mu\,  \frac{2n_1^r \Del_1}{R}\, \frac{1 - \frac{7M}{R} + \frac{18M^2}{R^2}}{\lh 5 - \frac{18M}{R} \rh
 \lh 1 - \frac{2M}{R} \rh},  }\\
 \\
\dsp{ m_2^{\vf} = \mu\, \frac{n_1^{r\,2}}{4 R}\, \frac{5 - \frac{32M}{R}}{1 - \frac{6M}{R}}; }
\ea
\label{2.5.8}
\ee
d.\ and finally, for the angular frequency:
\be
\og_0 = \sqrt{\frac{M}{R^3} \frac{1 - \frac{6M}{R}}{1 - \frac{3M}{R}}}, \hs{1}
\og_1 = - \og_0\, \frac{3 \Del_1}{2R}\, \frac{1 - \frac{10M}{R} + \frac{18M^2}{R^2}}{\lh 1 - \frac{3M}{R} \rh
 \lh 1 - \frac{6M}{R} \rh}. 
\label{2.5.9}
\ee
For fixed mass $M$, the undetermined constants in these expressions are $(R, \sg, \Del_1, \Del_2, ...)$. In every order there 
is one additional parameter to be fixed by initial conditions: $R$ to fix the circular reference orbit, $\sg$ to fix the initial distance 
from the circular orbit (here: the periastron), and the shape parameters $\Del_n$ determining the apastron, periastron advance, 
and other orbital characteristics.  

By construction the orbits (\ref{2.3}) imply the existence of the constants of motion $\ve$ and $\ell$ defined in eq.\ (\ref{2.2}): 
\[
\ve = \lh 1 - \frac{2M}{r} \rh \frac{dt}{d\tau}, \hs{2} \ell = r^2 \frac{d\vf}{d\tau},
\]
such that
\be
\frac{d\ve}{d\tau} = 0, \hs{2} \frac{d\ell}{d\tau} = 0.
\label{2.6}
\ee
In terms of the expressions (\ref{2.3}), (\ref{2.5.1}) these identities indeed hold order by order in $\sg$ for all allowed values of
$(R, \sg, \Del_n)$, the restrictions being $R \geq 6M$ and $r \geq 2M$. The values of $\ve$ and $\ell$ are then given to order
$\sg^2$ by
\be
\ve = \ve_0 + \sg \ve_1 + \frac{1}{2}\, \sg^2 \ve_2 + ..., \hs{2}
\ell = \ell_0 + \sg \ell_1 + \frac{1}{2}\, \sg^2 \ell_2 + ...,
\label{2.6.1}
\ee
with
\be
\ba{lll}
\ve_0 & = & \dsp{ \frac{1 - \frac{2M}{R}}{\sqrt{ 1 - \frac{3M}{R} }}, \hs{2}
\ve_1 = \frac{1}{2} \frac{\og_0^2 R \Del_1}{\sqrt{ 1 - \frac{3M}{R} }}, }\\
 & & \\
\ve_2 & = & \dsp{ \frac{1}{2} \frac{\og_0^2 R \Del_2}{\sqrt{1 - \frac{3M}{R}}} - \frac{M \lh n_1^r \rh^2}{2R^3}\, 
 \frac{1 - \frac{9M}{R} + \frac{6M^2}{R^2}}{\lh 1 - \frac{2M}{R} \rh \lh 1 - \frac{3M}{R} \rh^{3/2}} }\\
 & & \\
 & & \dsp{ - \frac{M \Del_1^2}{4R^3}\, \frac{22 - \frac{288M}{R} + \frac{1279M^2}{R^2} - \frac{1836 M^3}{R^3} + \frac{972 M^4}{R^4}
  }{\lh 1 - \frac{3M}{R} \rh^{5/2} \lh 1 - \frac{6M}{R} \rh}; }
\ea
\label{2.6.2}
\ee
and
\be
\ba{lll}
\ell_0 & = & \dsp{ \sqrt{\frac{MR}{1 - \frac{3M}{R}}}, \hs{2} 
\ell_1 = \frac{1}{2} \sqrt{\frac{R}{M}} \frac{\og_0^2 R^2 \Del_1}{\sqrt{1 - \frac{3M}{R}}}, }\\
 & & \\
\ell_2 & = & \dsp{ \frac{1}{2} \sqrt{\frac{R}{M}} \frac{\og_0^2 R^2 \Del_2}{\sqrt{1 - \frac{3M}{R}}} 
 - \frac{3 \lh n_1^r \rh^2}{2R} \sqrt{\frac{M}{R}} \frac{1 - \frac{7M}{R}}{\lh 1 - \frac{3M}{R} \rh^{3/2}} }\\
 & & \\
 & & \dsp{ - \frac{\Del_1^2}{4R} \sqrt{\frac{M}{R}} \frac{19 - \frac{243M}{R} + \frac{963M^2}{R^2} - \frac{1512M^3}{R^3}
 + \frac{972M^4}{R^4}}{\lh 1- \frac{3M}{R} \rh^{5/2} \lh 1 - \frac{6M}{R} \rh}. }
\ea
\label{2.6.3}
\ee
Given values of the parameters $(R, \sg, \Del_1, ..., \Del_n)$ up to some order $n$, we obtain a curve approximating 
a geodesic up to and including terms of order $\sg^n$, and the corresponding values of the constants of motion $\ve$ 
and $\ell$. Going to order $n+1$  whilst keeping the values of the parameters $(R, \sg, \Del_1, .., \Del_n)$ fixed then 
changes  the approximate values of $\ve$ and $\ell$, in a way depending on the choice of $\Del_{n+1}$. In the limit 
$n \rightarrow \infty$ the curve approaches an exact geodesic, but in general this geodesic is not characterized by 
the initial values of the constants of motion for the circular orbit. The large redundancy introduced by the infinite set 
of parameters $\Del_n$, which together determine the radius $a_0^r$ of the fundamental circular orbit,  is easily 
understood, as any circular orbit can be deformed continuously into any other periodic (bound) orbit, including other 
circular orbits. The redundancy can be lifted in a simple way by taking all $\Del_n = 0$, which reduces the expansion 
to the one constructed in ref.\ \ct{OriginalEpicycle}. Then $n_1^r$ is fixed by the first eq.\ (\ref{2.5.6}) and the values 
of the constants of motion take the values dictated by the choice of $R$ and $\sg$ in (\ref{2.6.2}) and (\ref{2.6.3}), 
which change order by order in the expansion. 

Alternatively, one can use the parameters $\Del_n$ to improve the approximation of a particular orbit at fixed order 
in $\sg$. For example, one can require the constants of motion $\ve$ and $\ell$ to have a fixed value at all orders 
in $\sg$ from the first order onwards. However, in general this can not be done by adjusting the single parameter 
$\Del_{n}$ at order $n$, and as a result the actual values of all parameters $(R, \sg, \Del_1, ..., \Del_n)$ will have 
to be readjusted order by order.

In practice, we prefer to use the freedom in the orbital parametrization to adjust all parameters so as to keep a 
growing number of orbital characteristics fixed order by order in the expansion: the orbital period at zeroth order, the 
radial co-ordinate of the periastron and the periastron advance at first order, the apastron co-ordinate at second 
order, etc., and to keep them fixed henceforth order by order in the expansion. Again, to this end the actual values of 
the parameters $(R, \sg, \Del_1, ..., \Del_n)$, and correspondingly the constants of motion $\ve$ and $\ell$, have to 
be readjusted at every order $n$. For the second-order approximation presented above, these conditions read 
explicitly\footnote{Our initial conditions are such that the test mass is at periastron at $\tau = 0$; therefore 
$a_1^r$ and $\sg$ are necessarily negative.} 
\be
r_{pa} = a_0^r + a_1^r + a_2^r + ..., \hs{2} r_{aa} = a_0^r - a_1^r + a_2^r + ...,
\label{2.7}
\ee
whilst the total proper-time period $\Del \tau$, observer time period $\Del t$ and angle $\Del \vf$ between periastra are
\be
\og \Del \tau = (\og_0 + \sg \og_1 + ...) \Del \tau = 2 \pi, 
\label{2.8}
\ee
and hence 
\be
\og \Del t = 2\pi a_0^t, \hs{2} 
\og \Del \vf = 2\pi a_0^{\vf}.
\label{2.9}
\ee
These four conditions determine the values of $(R, \sg, \Del_1, \Del_2)$ at second order. 

\section{Numerical results \label{num}}

The accuracy of our perturbation theory can be tested by comparison with the results of a purely numerical approach.
As shown in the appendix of ref.\ \ct{GKJWvH}, we can parametrize any bound orbit in terms of a quasi semi-major
axis $a$ and a quasi eccentricity $e$ such that
\be
r = \frac{a}{1 + e \cos y(\vf)}, \hs{2} \lh \frac{dy}{d\vf} \rh^2 = 1 - \frac{2M}{a} \lh 3 + e \cos y \rh.
\label{4.1}
\ee
The parameters $(a, e)$ are determined by the constants of motion $(\ve, \ell)$ via
\be
\ba{l}
\dsp{ \ve^2 = \lh 1 - \frac{2M}{a} \rh \lh 1 + \frac{\ell^2}{a^2} \rh + \frac{e^2 \ell^2}{a^2} \lh 1 - \frac{6M}{a} \rh, }\\
 \\
\dsp{ a^2 - \frac{a \ell^2}{M} + \ell^2 \lh 3 + e^2 \rh = 0. }
\ea
\label{4.2}
\ee
In this section we perform a detailed comparison of both the first- and second-order perturbation expressions
with a series of high-precision numerical solutions of the geodesic equation characterized by specific values
of $a$ and $e$. It is shown that the accuracy varies rather little with decreasing values of $a$, affirming that 
the epicycle approximations remain very accurate all the way up to the ISCO. 

As a typical example we consider an orbit with $a = 10M$ and $e = 0.1$. For this orbit the radial co-ordinate
of the periastron is $r_{pa} = 9.09091 M$, whilst the apastron is found at $r_{aa} = 11.1111 M$. Recall, that 
the ISCO is the circular orbit at $r = 6M$. Furthermore, the angular shift between consecutive periastra is
large: $\del \vf = \Del \vf - 2\pi = 3.6561$. This shows that for the orbit considered the effects of the curvature 
of space-time are large. In the following for convenience of numerical calculations we take $M = 10$
in some arbitrary units. In these units the proper time between the periastra is $\Del \tau = 2661.05$, and 
the energy and angular momentum per unit of mass are
\be
\ve = 0.956568, \hs{2} \ell = 37.8235. 
\label{4.3}
\ee
Now we construct the first-order epicycle approximation to this geodesic. From eqs.\ (\ref{2.3})-(\ref{2.5.9}) we get
\be
\ba{l}
\dsp{ r_{pa} = R + \sg \lh \Del_1 + n_1^r \rh = R + \sg \sqrt{1 - \frac{2M}{R}}, }\\
 \\
\dsp{ \Del \tau = \frac{2\pi}{\og_0} = 2\pi R\, \sqrt{ \frac{R}{M} \frac{1- \frac{3M}{R}}{1 - \frac{6M}{R}} }, }\\
 \\
\dsp{ \Del \vf = a_0^{\vf}\, \Del \tau = 
 \frac{2\pi}{\sqrt{ 1 - \frac{6M}{R} }} \lh 1 - \frac{3\sg \Del_1}{2R}
 \frac{1 - \frac{2M}{R}}{1 - \frac{3M}{R}} \rh. }
\ea
\label{4.4}
\ee
These three conditions completely determine $R$, $\Del_1$ and $\sg$, 
\be
R = 101.274, \hs{1} \Del_1 = 0.05047, \hs{1} \sg = - 11.5699.
\label{4.5}
\ee
From these results we can compute the other dependent parameters and constants of motion:
\be 
n_1^r = 0.845359, \hs{1} \og_0 = 0.00236, \hs{1} \ve = \ve_0 + \sg \ve_1 = 0.956418, \hs{1}
 \ell = \ell_0 + \sg \ell_1 = 37.8710.
\label{4.6}
\ee
Observe, that the first-order values of $\ve$ and $\ell$ are accurate to about one per mille. 
We thus find the following explicit first-order approximation to the orbit:
\be
\ba{lll}
t(\tau) & = & 1.19347\, \tau+ 11.998\, \sin (0.00236\, \tau), \\
 \\
r(\tau) & = & 100.690 - 9.78072\, \cos (0.00236\, \tau), \\
 \\
\vf(\tau) & = & 0.00373\, \tau + 0.3025\, \sin (0.00236\, \tau). 
\ea
\label{4.7}
\ee

\begin{figure}[!h]
\centering
  \subfloat{\includegraphics[scale=0.68]{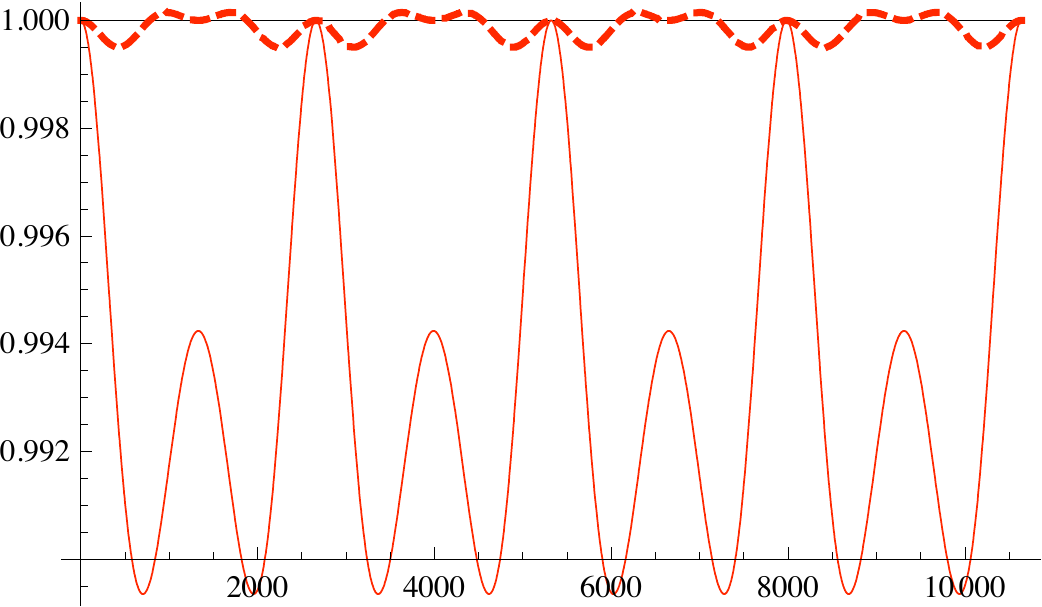}}
    \caption{\small The ratio $r(\tau)/r_{num}(\tau)$ in first-order approximation (solid curve) and in second-order 
    approximation (dashed curve).}
    \label{FigureOrbits1}
\end{figure}

\begin{figure}[!h]
\centering
  \subfloat{\includegraphics[scale=0.68]{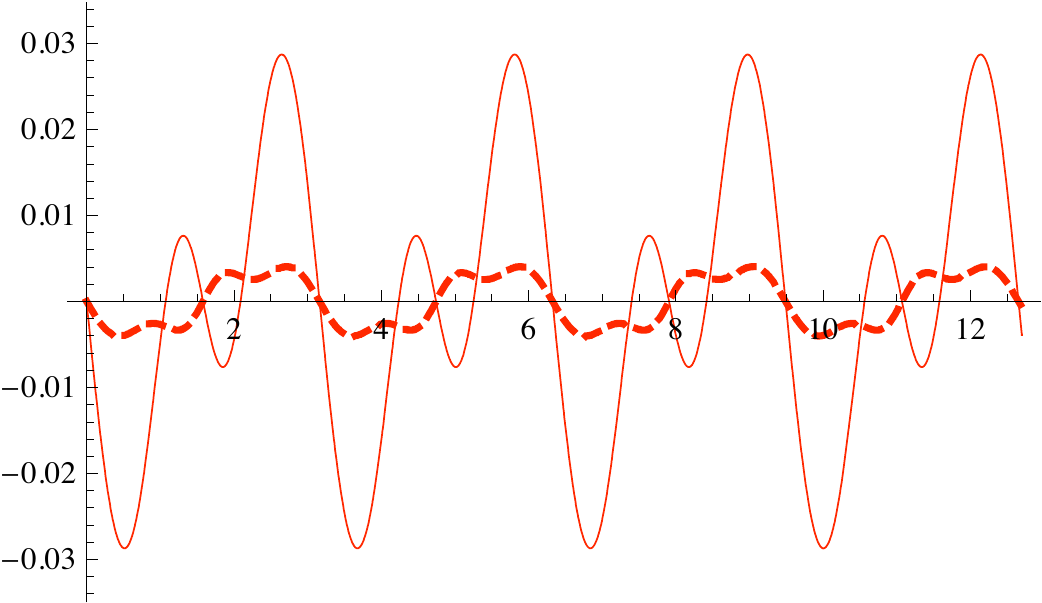}}
    \caption{\small The difference $\vf(\tau) - \vf_{num}(\tau)$ in radians, in first-order approximation (solid curve) 
    and in second-order approximation (dashed curve).}
    \label{FigureOrbits2}
\end{figure}

\nit
Next we turn to the second order approximation to the exact orbit. As boundary conditions we impose
\be
\ba{l}
\dsp{ r_{pa} = R + \sg \lh \Del_1 + n_1^r \rh + \frac{\sg^2}{2} \lh \Del_2 + n_2^r + m_2^r \rh, }\\
 \\
\dsp{ r_{aa} =  R + \sg \lh \Del_1 - n_1^r \rh + \frac{\sg^2}{2} \lh \Del_2 - n_2^r + m_2^r \rh, }\\
 \\
\dsp{ \Del \tau = \frac{2\pi}{\og_0 + \sg \og_1} \simeq \frac{2\pi}{\og_0} \lh 1 - \frac{\sg \og_1}{\og_0} \rh , }\\
 \\
\dsp{ \Del \vf = a_0^{\vf}\, \Del \tau = \frac{2\pi}{\og_0 R} \left[ \frac{\ell_0}{R} + \sg \lh R v_1^{\vf} - 
 \frac{\og_1 \ell_0}{\og_0 R} \rh + \frac{\sg^2 R}{2} \lh v_2^{\vf} - \frac{2 \og_1}{\og_0}\,  v_1^{\vf} \rh \right]. }
\ea
\label{4.8}
\ee
Given the values of periastron and apastron, and the periastron shift in angle and proper time as above, 
these equations determine $R$, $\Del_1$, $\Del_2$ and $\sg$; however, as the equations are quadratic in 
$\sg$, there are in general multiple solutions. We always choose the one with the smallest ratio of second-order to
first-order contributions to $\vf(\tau)$; in practice we find that this also implies fastest convergence for $r(\tau)$.
Using this criterion our second-order solution is characterized by 
\be
R = 100.046, \hs{1} \Del_1 = -0.12613, \hs{1} \Del_2 = 0.000228, \hs{1} \sg = -9.71137.
\label{4.9}
\ee
Observe, that the values of $R$, $\Del_1$ and $\sg$ differ from those at first order, as we are optimizing the fit  of
our approximation to a {\em given} orbit order-by-order in perturbation theory. For the other parameters the results 
(\ref{4.10}) imply 
\be
n_1^r = 1.02061, \hs{1} \og  = 0.00236, \hs{1} \ve = 0.95667, \hs{1} \ell = 37.8263.
\label{4.10}
\ee
Thus the explicit second-order solution reads
\be
\ba{lll}
t(\tau) & = & 1.19246\, \tau + 12.3903\, \sin (0.00236\, \tau) + 0.613326\, \sin (0.00472\, \tau), \\
 & & \\
r(\tau) & = & 101.378 - 10.101\, \cos (0.00236\, \tau) - 0.368364\, \cos (0.00472\, \tau), \\
 & & \\
\vf(\tau) & = & 0.0037351\, \tau + 0.312458\, \sin (0.00236\, \tau) + 0.0174545\, \sin (0.00472\, \tau).
\ea
\label{4.11}
\ee
To compare these results with the purely numerical solution in figs.\ 1 and 2 we plot the ratio of our perturbative
$r(\tau)$ to the numerical result $r_{num}(\tau)$, and the difference of our perturbative $\vf(\tau)$ and the numerical 
values $\vf_{num}(\tau)$. These figures show, that our approximations to the radial distance are accurate to better than 
1.1 \% at first order and to better than 0.05 \% at second order, whilst the accuracy of the azimuthal co-ordinate is 0.029 
radians at first order, and 0.004 radians at second order. 

\begin{table}
\begin{center} 
\begin{tabular}{ |  l || c | c | c | c | c |   r | } 
\hline
\multicolumn{1}{|c||}{$e=0.1$} & \multicolumn{2}{|c|}{First order epicycles} & \multicolumn{2}{|c|}{Second order epicycles} \\
\hline
$a$ & max. rel. diff. in $r$ & max. abs. diff. in $\varphi$ & max. rel. diff. in $r$ & max. abs. diff. in $\varphi$ \\ 
\hline
\hline $200$ & $1.05\%$ & 0.018 & $0.06 \%$ & $0.0015$ \\ 
\hline $150$ & $1.03\% $ & 0.020 & $0.06 \%$ & $0.0018$\\ 
\hline $100$ & $1.08 \%$ & 0.029 & $0.05 \%$ & $0.0040$\\ 
\hline $90$ & $1.09 \%$ & 0.024 & $ 0.05 \%$ & $0.0070$ \\ 
\hline $85$ & $1.04 \%$ & 0.042 & $0.04 \%$ & $0.012$ \\
\hline $80$ & $1.21 \%$ & 0.058 & $0.04 \%$ & $0.026$\\ 
\hline $75$ & $1.47 \%$ & 0.028 &$0.08 \% $ & $0.053$ \\ 
\hline $70$ & $1.52 \%$ & 0.040 & $0.12 \% $ & $0.012$\\ 
\hline $66$ & $2.20 \%$ & 0.022 & $0.24 \%$ & $0.020$\\ 
\hline
\end{tabular} 
\caption{\small Accuracy of epicycle approximation for orbits of eccentricity $e=0.1$ and various values of $a$, as listed 
in the first column. The second column presents the maximum relative differences between $r(\tau)$ at first order and its 
numerical counterpart; the third column presents the maximum absolute difference between the azimuth $\varphi(\tau)$ at 
first order and its numerical counterpart. The fourth  and fifth column present the same information for the second-order approximations.
\label{table:Table}}
\end {center}
\end{table}

To test the accuracy of our perturbation theory in the regime of strong curvature we have performed a series of 
similar calculations for values of $a$ ranging from $20 M$ to $6.6 M$, all with quasi eccentricity $e = 0.1$, such 
that the smallest orbit actually reaches the ISCO at $6M$ at its periastron. Again, for numerical purposes we have 
chosen to set the central mass equal to $M = 10$ in arbitrary units. The comparison with numerical calculations
of the corresponding orbits is summarized in table 1.

\np
\section{Gravitational waves \label{gw}}

Having in hand the perturbative solution of the geodesic equations for test masses in a Schwarzschild background, 
we can use them to compute the gravitational radiation emitted by extreme mass ratio binary systems in which the 
larger mass is non-rotating, at least in the limit in which radiation reaction effects can be ignored. The two linearized 
fluctuation modes of the Schwarzschild geometry are described by the Regge-Wheeler and Zerilli-Moncrief equations 
\ct{RW,zerilli,moncrief}. A test mass moving in a Schwarzschild background acts as a source for such fluctuations, 
giving rise to a specific type of source terms in the fluctuation equations. 

In a multipole expansion the fluctuation equations for the amplitudes $\Psi^{lm}_X$, $X = (RW, ZM)$ denoting the 
Regge-Wheeler or Zerilli-Moncrief modes, take the form
\be 
\lh \Box_S - V^l_X(r) \rh \Psi_X^{lm}(t,r) =  S^{lm}_X(t,r), 
\label{gw.1}
\ee
where the radial d'Alembert operator is given by 
\be
\Box_S = - \der_t^2 +  f \der_r f \der_r, \hs{2}
f(r) = 1 - \frac{2M}{r}.
\label{gw.2}
\ee
The potentials $V_X^l(r)$ depend on the radial co-ordinate $r$ and the mode-index $l$:
\be
\ba{lll}
V^l_{RW} & = & \dsp{ \frac{f(r)}{r^2} \lh l (l+1) - \frac{6M}{r} \rh,  }\\
 & & \\
V^l_{ZM} & = & \dsp{ \frac{f(r)}{\lh r + \frac{6M}{\lb} \rh^2} \lh l (l+1) + \frac{6M}{r} + \frac{36M^2}{\lb r^2} + \frac{72M^3}{\lb^2 r^3} \rh, } 
\ea
\label{gw.3}
\ee
where $\lb = \lh l + 2 \rh \lh l-1 \rh$. Using the results of section \ref{GeodesicDeviationMethod} the source terms can be 
computed from the energy-momentum tensor for a point mass $\mu$
\be
T_{\mu\nu} = \mu\, \int \frac{1}{\sqrt{-g}}\, u_{\mu} u_{\nu}\, \del^4\lh x - x_p(\tau) \rh d\tau,
\label{gw.3.1}
\ee
to take the form 
\begin{equation}
S^{lm}_X = G^{lm}_X(r_p(t))\, \delta(r-r_p(t)) + F^{lm}_X(r,r_p(t))\, \delta^{\prime}(r-r_p(t)), 
\label{gw.4}
\end{equation}
in which 
\begin{equation}
\ba{lll}
F^{lm}_{RW}(r,r_p(t)) &=& \dsp{ -16  \pi \mu\, \frac{(l-2)!}{(l+2)!}\, \frac{r_p^2 f^2(r)}{r} \frac{(u^\varphi)^2}{u^t} W^{*lm}_{\varphi \varphi}, }\\
 \\
G^{lm}_{RW}(r,r_p(t)) &=& \dsp{ 32 \pi \mu\,  \frac{(l-2)!}{(l+2)!}\, f(r_p) \left(1-\frac{3M}{r_p}\right)\, 
 \frac{(u^\varphi)^2}{u^t} W^{*lm}_{\varphi \varphi} + \frac{16 \pi \mu}{l(l+1)}\, \frac{f(r_p)}{r_p}\,  
 \frac{u^r u^\varphi}{u^t}X^{*lm}_\varphi, }
\ea
\label{gw.5}
\end{equation}
and 
\begin{eqnarray}
F^{lm}_{ZM}(r,r_p(t)) &=& \frac{32\pi \mu}{(\lambda+2)\lh \lb + \frac{6M}{r_p} \rh}\, \frac{ r^2 f(r) }{r_p^2} \left( f^2(r) u^t  
 - \frac{(u^r)^2}{u^t} \right) Y^{*lm} \nonumber \\
G^{lm}_{ZM}(r,r_p(t)) &=& \frac{16 \pi \mu}{r_p}\, \frac{\lh \lb + \frac{6M}{r_p} - 2f(r_p) \rh}{(\lambda+2) \lh \lb + \frac{6M}{r_p}  \rh}\, 
 \frac{(u^r)^2}{u^t} Y^{*lm} + \frac{32 \pi \mu\, r_p f^2(r_p)}{(\lambda+2) \lh \lb + \frac{6M}{r_p} \rh}\,  \frac{(u^\varphi)^2}{u^t} U^{*lm}_{\varphi \varphi} \nonumber \\
& & -\, \frac{16 \pi \mu  f^2(r_p)}{r_p}\, \frac{ \left( \lambda(\lambda-2) +
 \lh 2\lambda-9 \rh \frac{4M}{r_p} + \frac{60 M^2}{r_p^2} \right)}{(\lambda+2) \lh \lb + \frac{6M}{r_p} \rh^2}\, u^t\, Y^{*lm} \nonumber \\
& & +\, \frac{64 \pi \mu}{l(l+1)} \frac{f(r_p)}{\lh \lb + \frac{6M}{r_p} \rh}\, \frac{u^r u^\varphi}{u^t}\, Z^{*lm}_{\varphi} 
 - 32 \pi \mu r_p\, \frac{(l-2)!}{(l+2)!}\,  \frac{(u^\varphi)^2}{u^t}\, V^{*lm}_{\varphi \varphi}.
\label{gw.6}
\end{eqnarray}
Here $Y^{lm}$, $(X_A^{lm}, Z_A^{lm})$ and $(U_{AB}^{lm}, V_{AB}^{lm}, W_{AB}^{lm})$ represent standard scalar, 
vector and tensor harmonics \ct{chandra,Martel}, whilst the four-velocity components $u^{\mu}$ are to be taken from sect.\ 
\ref{GeodesicDeviationMethod}.  Details will be given in \ct{koekoek}.

The fluctuation equations (\ref{gw.1}) have been worked out and solved in a fully numerical approach for the source terms 
and fluctuations in refs.\ \ct{RW}-\ct{Martel}. We have numerically solved the equations for the asymptotic gravitational-wave 
amplitude at distance $r$:
\be
h_+(t) - i h_{\times}(t) = \frac{1}{r} \sum_{lm} \lh \Psi_{ZM}^{lm}(t) -  2i \int_{-\infty}^t \Psi^{lm}_{RW}(t^{\prime}) dt^{\prime} \rh 
 V_{AB}^{lm}\, \bar{m}^A \bar{m}^B,
\label{gw.7}
\ee
starting from the fluctuation equations in full analytic form, using the algorithm of Lousto and Price \ct{lousto-price},  
for the second-order approximation to the orbit discussed above, eqs.\ (\ref{4.9})-(\ref{4.11}).

From the amplitudes one can calculate the power emitted in terms of gravitational waves from the Regge-Wheeler and 
Zerilli-Moncrief functions by evaluating the expression \ct{Martel2, Martel}
\be
P = \frac{1}{64\pi} \sum_{lm} \frac{(l+2)!}{(l-2)!} \lh \left| \dot{\Psi}_{ZM}^{lm} \right|^2 + 4 \left|\Psi_{RW}^{lm} \right|^2 \rh,
\label{gw.8}
\ee
where the overdot denotes a derivative w.r.t.\ co-ordinate time $t$. Also, the angular momentum per unit time emitted by gravitational waves follows as
\begin{equation}
\frac{dL}{dt} = \frac{i}{128 \pi} \sum_{lm}  \frac{m (l+2)!}{(l-2)!} \left( \dot{\Psi}^{lm}_{ZM} \Psi^{*lm}_{ZM}+4 \Psi_{RW}^{*lm} \int_{-\infty}^{t} \Psi^{*}_{RW}(t') dt'  \right) + c.c .
\label{ChangeAngularMomentum}
\end{equation}
in which $c.c.$ stand for the complex conjugate.

\begin{table}[!h]
\begin{center} 
\begin{tabular}{ |  l || c | c | c | c |   r | } 
\hline
$a$ & $\langle P \rangle_{epicycle}$ & $\langle P \rangle_{PM}$  & rel. diff.\\ 
\hline
\hline $200$ & $2.033\ 10^{-6}$ & $2.033\ 10^{-6}$ & $ + 0.2 \%$\\ 
\hline $150$ & $8.139\ 10^{-6}$ & $8.555\ 10^{-6}$ & $ - 5.1 \%$\\ 
\hline $100$ & $6.294\ 10^{-5}$ & $6.496\ 10^{-5}$ & $ - 3.2 \%$\\ 
\hline $90$  & $1.083\ 10^{-4}$ & $1.100\ 10^{-4}$ & $ - 1.6 \%$\\ 
\hline $85$ & $1.475\ 10^{-4}$ & $1.464\ 10^{-4}$ & $ -0.5 \%$\\ 
\hline $80$ & $1.991\ 10^{-4}$ & $1.983\ 10^{-4}$ & $ +0.4 \%$\\ 
\hline $75$ & $2.944\ 10^{-4}$ & $2.738\ 10^{-4}$ & $ +7.6 \%$\\ 
\hline $70$ & $4.209\ 10^{-4}$ & $3.865\ 10^{-4}$ & $ + 8.9 \%$\\ 
\hline $66$ & $5.869\ 10^{-4}$ & $5.187\ 10^{-4}$ & $ + 13 \%$\\ 
\hline
\end{tabular} 
\caption{\small Average power $\langle P \rangle$ emitted in gravitational waves by a system of a star of mass $\mu$ 
in a bound orbit of eccentricity $e=0.1$ around a black hole of mass $M=10$ for various values of $a$, as listed in the 
first column. The second column presents the power in the second order epicycle approximation, whereas the third 
column presents the power as calculated by the Peters-Mathews equation. The third column presents the relative 
difference in percents. All powers are computed to a numerical accuracy of $0.1\%$ and are stated in units $M^2/\mu^2$. 
\label{table:Table02}}
\end {center}
\end{table}

We have performed the computation of the gravitational-wave power for the full series of orbits listed in table 
\ref{table:Table}, which presents the power averaged over an orbital revolution. The results in units of the mass ratio 
$(M/\mu)^2$ are presented in the second column of table \ref{table:Table02}. They can be compared to the power as 
calculated by the Peters-Mathews equation in the Newtonian approximation \ct{peters-mathews} in the third column. 
To reach the required accuracy it is sufficient to compute the modes up to $l =4$, $m = 4$ and add them as indicated 
in eqs.\ (\ref{gw.7}) - (\ref{ChangeAngularMomentum}); 
higher-order contributions do not change the results. In fact, the main contribution comes from the modes with 
$l = 2$, as is to be expected from the quadrupole nature of free gravitational waves. The numerical accuracy 
has furthermore been checked by requiring the amplitude and power to remain the same upon reducing the 
grid scale by a factor two. 

\begin{figure}[!h]
\centering
  \subfloat{\includegraphics[scale=0.22]{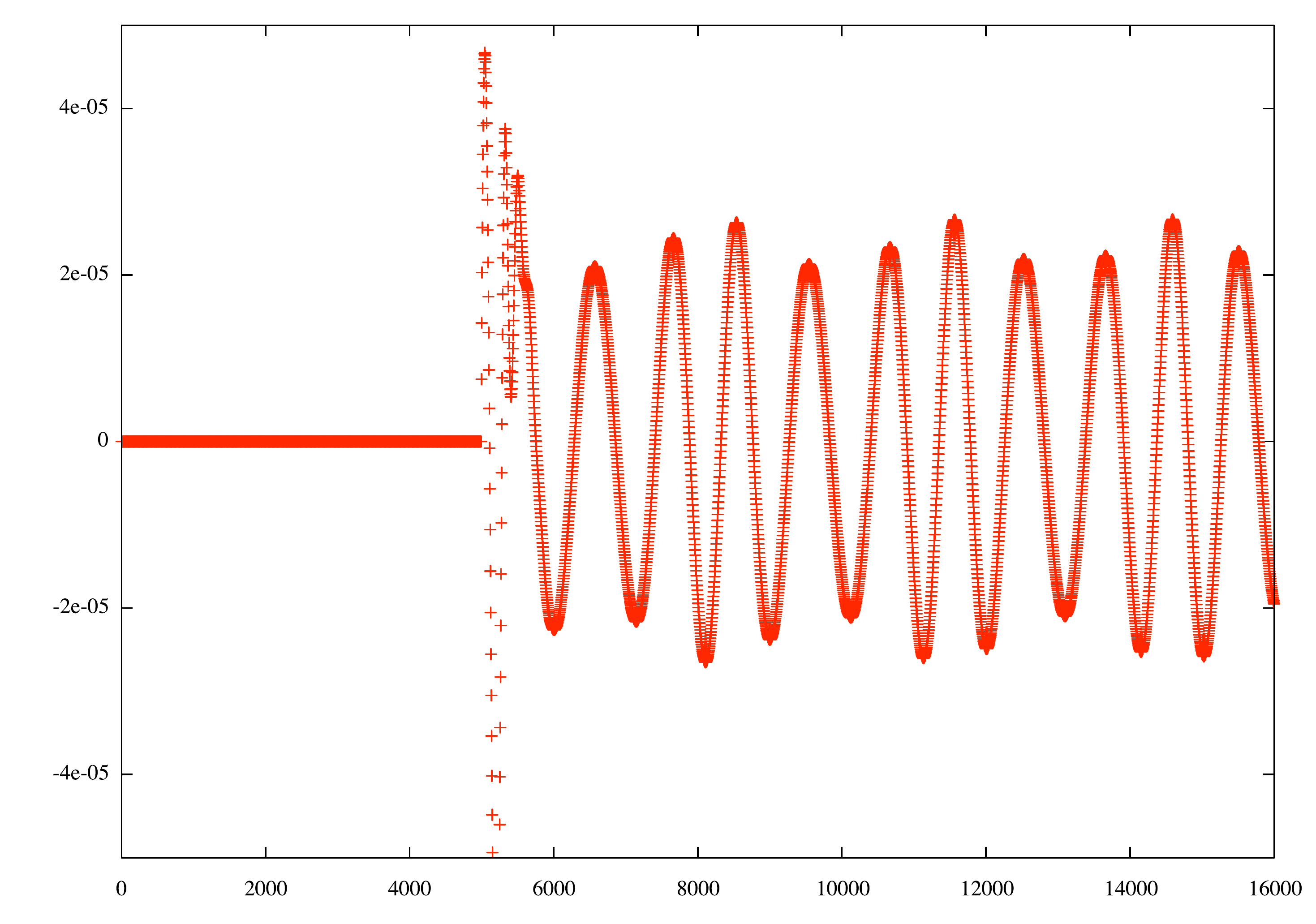} \hs{1} \includegraphics[scale=0.22]{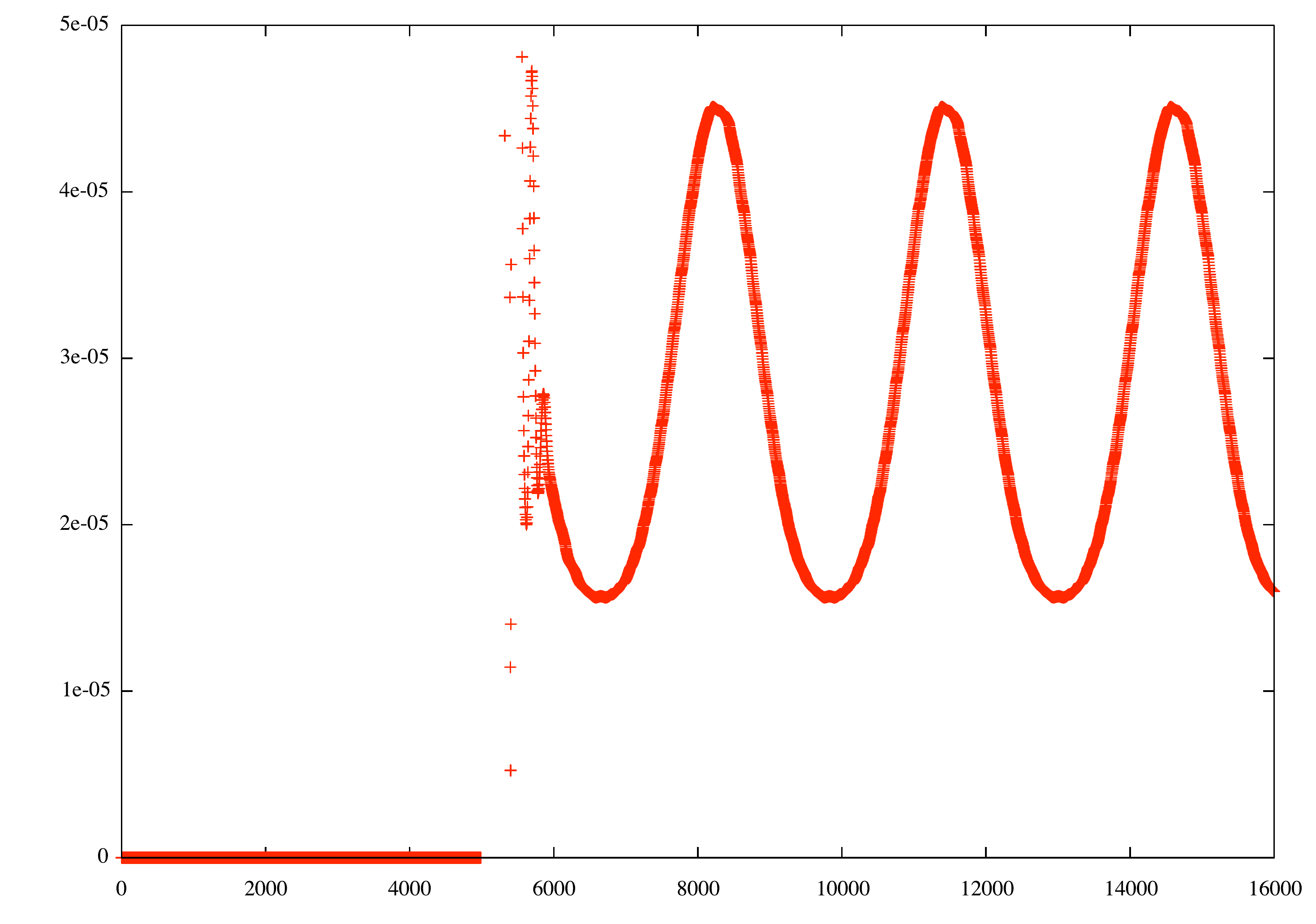}}
    \caption{\small Real part of the $ l = 2$, $m=2$ Zerilli-Moncrief function (left) and power emitted by this mode (right)
    as a function of time $t$, for an observer located at distance $r = 500M$.}
    \label{ZMRetime}
\end{figure}

\begin{figure}[!h]
\centering
  \subfloat{\includegraphics[scale=0.22]{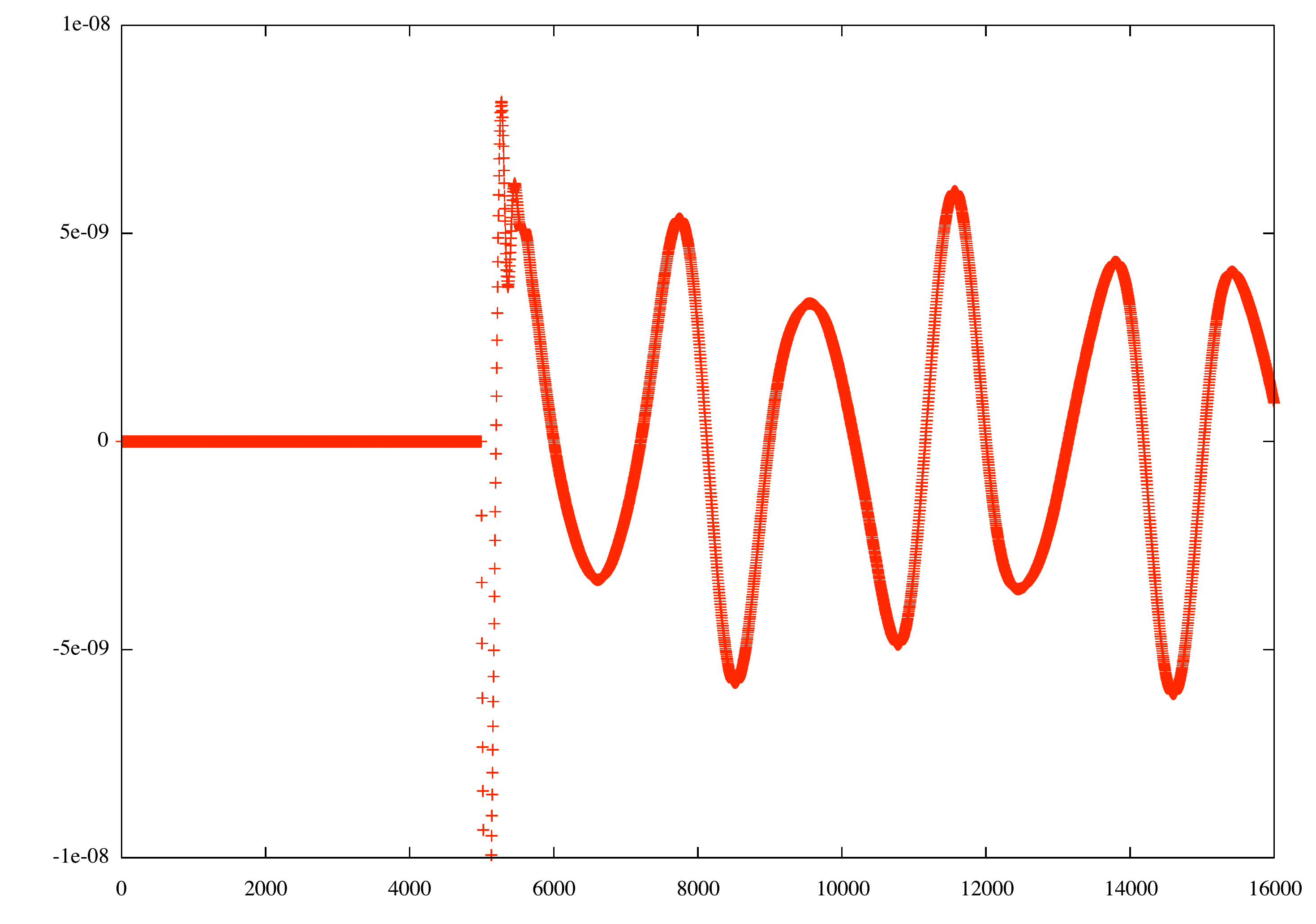} \hs{1} \includegraphics[scale=0.22]{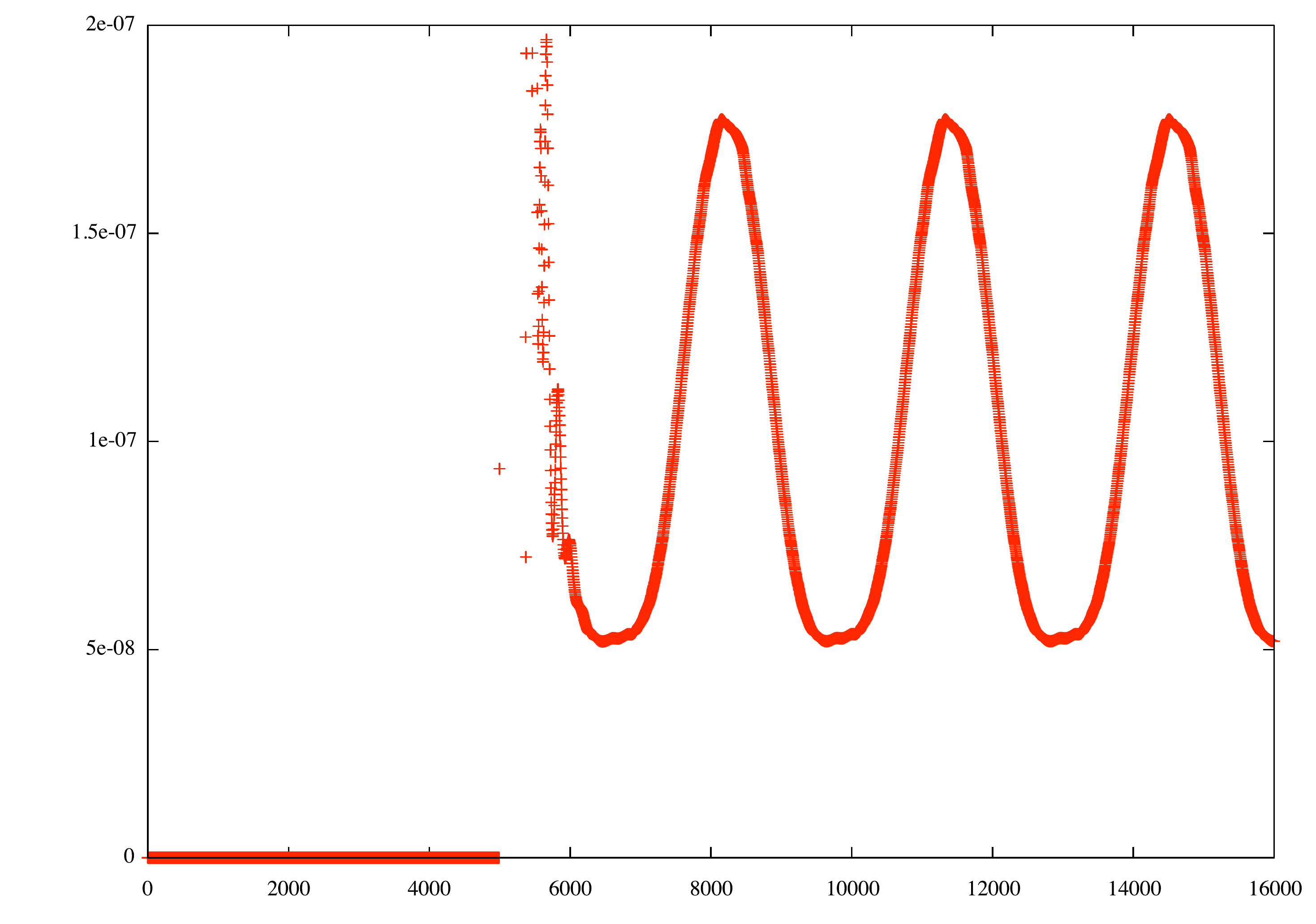}}
    \caption{\small Real part of the $ l = 2$, $m=1$ Regge-Wheeler function (left) and power emitted by this mode (right)
    as a function of time $t$, for an observer located at distance $r = 500M$.}
    \label{RWRetime}
\end{figure}

As an example, the $l=2$ modes are shown for the reference orbit with $a = 10 M$ and $e = 0.1$ in figs.\ 
\ref{ZMRetime} and \ref{RWRetime}. The average power emitted in this orbit in the second-order epicycle approximation 
is 
\be
\langle P \rangle = 6.294 \times 10^{-5},
\label{gw.9}
\ee
which differs a mere 0.3 \% from the fully numerical result \ct{Fujita, HopperEvans} 
\be
\langle P \rangle = 6.318 \times 10^{-5}.
\label{gw.10}
\ee

Finally, we proceed to calculate the emitted power and angular momentum per unit time for a series of eccentric orbits that 
are adiabatically related to each other by the emission of gravitational radiation. That is, we calculate the average power 
and angular momentum emitted by gravitational waves for each given eccentric orbit, and use these numbers to update 
the values of $\varepsilon$ and $\ell$ in a discrete step; the newly found values for $\varepsilon$ and $\ell$ then correspond 
to the next orbit in the series. The discrete step is chosen as follows: the next orbit will always be chosen to be the one that 
has a value $\ell$ that is $1\%$ smaller than that of the current orbit. In this way, we find in practice that successive orbits 
have a percentual change in periastra of less than $4\%$, which justifies the adiabatic approximation. The results are 
shown in Table \ref{table:Inspiral}. As before, the size of the grid was chosen such that the values for $<P>$ and $<\dot{L}>$ 
do not change more than at the $0.1\%$ level when taking a grid size a factor of two smaller still. 

\begin{table}[h]
\begin{center} 
\begin{tabular}{ |  c | c || c | c | c | c | c | c |  c | } 
\hline
 $\quad a$ & $e $ & $ <P> $ & $ <\dot{L}>$ &$\Del a/a$ & $\Del e/e$ & $\Delta t$ & max.rel.diff. $r$& max.abs.diff. $\varphi$ \\ 
\hline
\hline 120.0 & 0.1500 & 2.536 & 0.9961 & n.a. & n.a. & n.a. &0.20\% & 0.0085 \\ 
\hline 116.4 & 0.1447 & 2.951 & 1.125 & 3.1\% & 3.7\% & 4.021 & 0.18\% & 0.0080 \\ 
\hline 112.8 & 0.1410 & 3.451 & 1.258 & 3.2\% & 2.6\% & 3.526 & 0.16\% & 0.0080 \\ 
\hline 109.3  & 0.1378 & 4.051 & 1.410 & 3.2\% & 2.4\% & 3.121 & 0.15\% & 0.0081 \\ 
\hline 105.8  & 0.1349 & 4.785 & 1.589 & 3.3\% & 2.2\% & 2.756 & 0.14\% & 0.0083 \\ 
\hline 102.3   & 0.1322 & 5.683 & 1.794 & 3.4\% & 2.0\% & 2.423 & 0.13\% & 0.0087 \\ 
\hline 98.7   & 0.1300 & 6.774 & 2.039 & 3.6\% & 1.8\% & 2.123 & 0.12\% & 0.0094 \\ 
\hline 95.2   & 0.1287 & 8.193 & 2.326 & 3.7\% & 1.0\% & 1.850 & 0.11\% & 0.011 \\ 
\hline 91.6   & 0.1273 & 9.853 & 2.661 & 3.9\% & 1.1\% & 1.604 &0.11\% & 0.013 \\ 
\hline
\end{tabular}
\caption{\small A series of eccentric orbits parametrized by $a$ and $e$, as listed in the first two columns. The orbits are related 
by the emission of gravitational waves in an adiabatic way, as explained in the main text. The average power emitted is given in 
the third column in units $10^{-15} \left( M/\mu \right)^2$, and the average angular momentum emitted per unit time is given in the 
fourth column in units $10^{-12} \left( M/\mu^2 \right)$. The fifth and sixth columns list the percentual change in $a$ and $e$
compared to the next larger orbit, showing explicitly the inspiral and circularization due to the emission of gravitational waves. 
The seventh column presents the Schwarzschild time taken to make the discrete step from the previous orbit to the current, in units $10^{11}$ seconds. The eighth column presents the maximum relative difference between the second-order epicycle radial orbital function $r(\tau)$ 
and its numerical counterpart; the last column presents the maximum absolute difference between the second-order epicycle 
angular orbital function $\varphi(\tau)$ and its numerical counterpart. The mass of the black hole and that of the companion star 
are, in arbitrary units, set to $M=10$ and $\mu = 10^{-4}$. \label{table:Inspiral}}
\end {center}
\end{table}

As can be seen, under influence of the emission of gravitational waves the eccentricity of the orbits decreases, and the 
orbits become more circular after every discrete step. As a result the epicycle expansion is expected to become 
increasingly accurate for successive orbits, and indeed this is seen to be the case: the radial orbital function $r$ becomes 
almost twice as accurate in the course of the orbits considered. In contrast, the precision of the angular coordinate $\varphi$ 
does not improve, but remains more or less fixed around the value of $0.008$ radians. This conforms to expectations, 
as in the second order epicycle expansion we use two of the four boundary conditions to fix he time and angular shift 
between successive periastra, leaving only two boundary conditions to fit the orbital functions $r$ and $\varphi$. In the 
calculation presented, these remaining two boundary conditions were used to fix the values of the radial positions of the 
periastron and apastron, which in practice also makes the orbital function $\varphi$ very accurate, but without causing
the accuracy to {\em increase} with decreasing eccentricity. 

\section{Discussion and conclusions} 

Comparing the results of the epicycle approximation with the Peters-Mathews calculations as collected in table 
\ref{table:Table02}, one observes that both at large distances and close to the ISCO the power computed by our 
relativistic procedure exceeds that of the newtonian approximation, whilst in contrast for intermediate distances 
the relativistic power is lower. This can be understood by the opposite effect of two factors: on the one hand the 
precession of the periastron shows that the orbital velocity and acceleration in the relativistic orbit is higher than 
that in the corresponding Kepler orbit; on the other hand in the relativistic computation the redshift of the gravitational 
waves lowers the power emitted as measured by a distant observer. 

Comparison with purely numerical calculations shows that for the series of orbits considered the second-order 
epicycle approximation is very accurate, at the level of one part in a thousand. The agreement is less for orbits 
with larger eccentricity; for $e = 0.2$ the agreement is still very good for large orbits such as $a= 20 M$,
but closer to the ISCO the accuracy becomes of the order of 1-5\%. To improve on this it is necessary to 
include the third-order epicycle contribution. Indeed, the source of deviations is the dependence on the azimuth 
angle $\vf(\tau)$, which appears as argument in the tensor spherical harmonics. The angular velocity can be
improved significantly without compromizing the radial accuracy only by including the third-order epicycle terms
\ct{GKJWvH}. However, as the emission of gravitational radiation tends to lead to significant loss of angular 
momentum, thus decreasing the eccentricity in the last stage of inspiral, in practice the second-order epicycle
approximation leads mostly to very acceptable results. 
\vs{2}

\nit
{\bf Acknowledgments} \\
This work is part of the research programme of the Foundation for Fundamental Research on Matter (FOM), which is part of the Netherlands Organisation for Scientific Research (NWO). We would like to thank Paul Zevenbergen for doing valuable preliminary work that made the numerical code possible.

\np

\end{document}